\documentclass[aps,prl,10pt,twocolumn,superscriptaddress,showpacs]{revtex4-1}
\usepackage{titlesec}
\usepackage{hyperref}
\usepackage{graphicx}
\usepackage{amsfonts,amsmath,amssymb,bm,bbm}
\usepackage{color}

\hypersetup{
    pdfnewwindow=true,      
    colorlinks=true,       
    linkcolor=blue,          
    citecolor=blue,        
    filecolor=blue,      
    urlcolor=blue        
}

\titleformat{\section}[runin]
  {\normalfont\bfseries}{\thesection}{1em}{}[:]
\titlespacing*{\section}{0cm}{2em}{1em}
\titleformat{\subsection}[runin]
  {\normalfont\itshape}{\thesubsection}{1em}{}[:]
\titlespacing*{\subsection}{1cm}{2em}{1em}

\begin{document}

\title{\mbox{\hspace{-0.25cm} Dark Matter Velocity Spectroscopy}}
\author{Eric G. Speckhard}
\affiliation{Center for Cosmology and AstroParticle Physics (CCAPP), Ohio State University, Columbus, OH 43210}
\affiliation{Department of Physics, Ohio State University, Columbus, OH 43210}

\author{Kenny C. Y. Ng}
\affiliation{Center for Cosmology and AstroParticle Physics (CCAPP), Ohio State University, Columbus, OH 43210}
\affiliation{Department of Physics, Ohio State University, Columbus, OH 43210}

\author{John F. Beacom}
\affiliation{Center for Cosmology and AstroParticle Physics (CCAPP), Ohio State University, Columbus, OH 43210}
\affiliation{Department of Physics, Ohio State University, Columbus, OH 43210}
\affiliation{Department of Astronomy, Ohio State University, Columbus, OH 43210}

\author{Ranjan Laha}
\affiliation{Kavli Institute for Particle Astrophysics and Cosmology (KIPAC),\\ Department of Physics, Stanford University, Stanford, CA 94035\\
SLAC National Accelerator Laboratory, Menlo Park, CA 94025 \\
{\tt speckhard.1@osu.edu, ng.199@osu.edu, beacom.7@osu.edu, rlaha@stanford.edu}}

\date{July 31, 2015}

\begin{abstract}
  Dark matter decays or annihilations that produce line-like spectra may be smoking-gun signals. However, even such distinctive signatures can be mimicked by astrophysical or instrumental causes. We show that velocity spectroscopy---the measurement of energy shifts induced by relative motion of source and observer---can separate these three causes with minimal theoretical uncertainties. The principal obstacle has been energy resolution, but upcoming experiments will reach the required $0.1\%$ level. As an example, we show that the imminent Astro-H mission can use Milky Way observations to separate possible causes of the 3.5-keV line. We discuss other applications.
\end{abstract}


\maketitle

\section{Introduction}
\label{sec:Introduction}

	What is the dark matter? Identification depends upon more than just observation of its bulk gravitational effects; distinct particle signatures are needed. Backgrounds make it difficult to pick out these signals, which are constrained to be faint. Among possible decay or annihilation signals, those with sharp spectral features, such as a line, are especially valuable.
	
	Given that the stakes and difficulties are so profound, even such a ``smoking-gun'' signal may not be conclusive. A line could have other causes: astrophysical (baryonic) emission or detector backgrounds (or response effects). For example, the cause of the recently discovered 3.5-keV line is disputed~\cite{Bulbul:2014sua,Boyarsky:2014jta,Riemer-Sorensen:2014yda,Boyarsky:2014ska,Anderson:2014tza,Malyshev:2014xqa,Jeltema:2014qfa,Urban:2014yda}. This problem is more general~\cite{Loewenstein:2009cm,Prokhorov:2010us,Weniger:2012tx,Finkbeiner:2012ez,2012arXiv1206.1616S,Aharonian:2012cs,Tempel:2012ey,Ackermann:2013uma,Weniger:2013tza,Ackermann:2015lka} and will surely arise again. We need better evidence than just a smoking gun---we need to see it in motion. 

\vspace{-0.5 cm}
\section{Premise and Motivation}

\begin{figure}[t]
\centering
\includegraphics[width=\columnwidth]{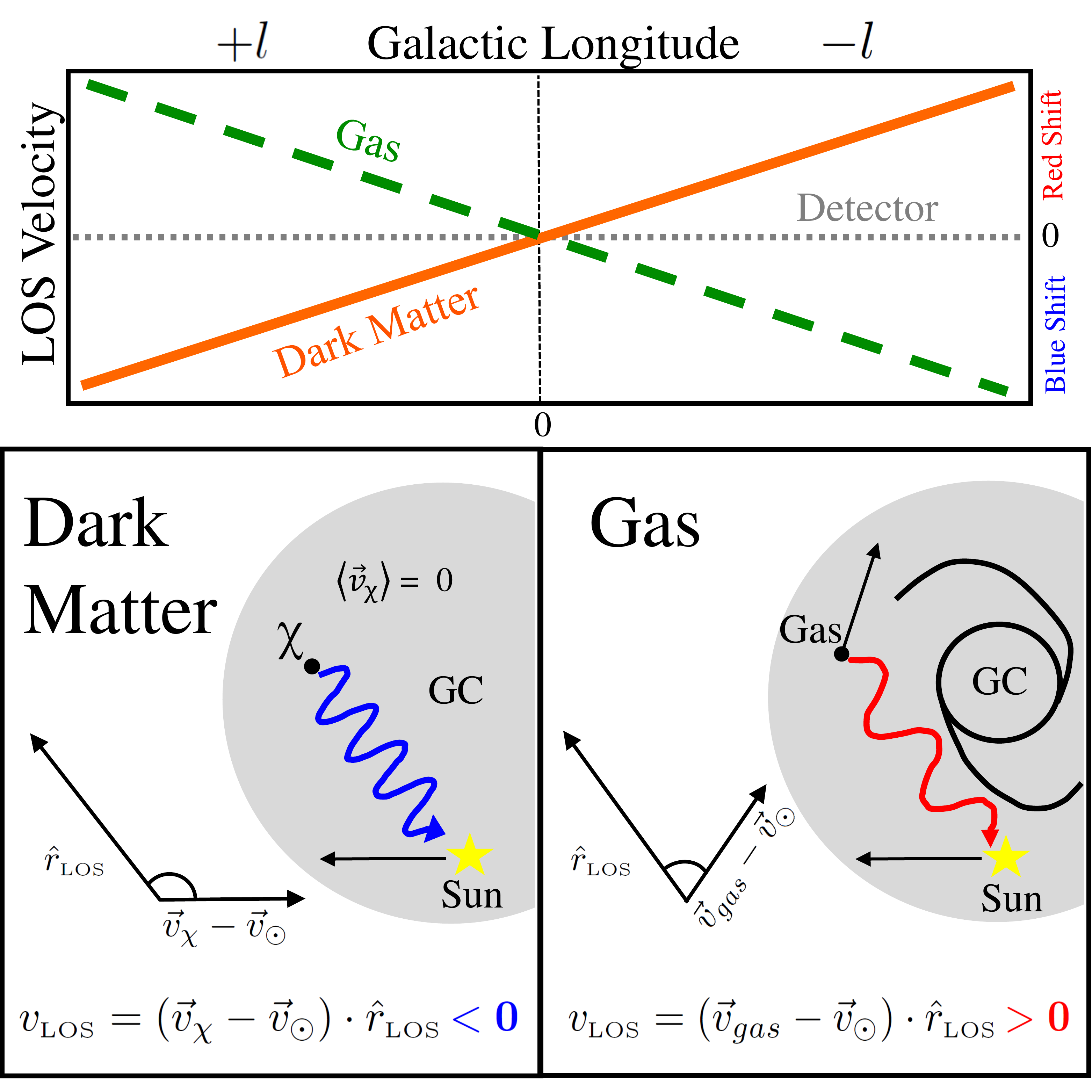}
\caption{\textbf{Top:} How DM, astrophysical, and detector lines shift with Galactic longitude is starkly different. \textbf{Bottom:} For DM signals at positive longitude, our motion through the non-rotating DM halo yields a negative LOS velocity and thus a blue shift. In contrast, for astrophysical lines (e.g., from gas), co-rotation in the disk leads to a positive LOS velocity and thus a red shift. These signs reverse at negative longitude. Detector lines have zero shift.} 
\label{fig:toon}
\vspace{-0.25 cm}
\end{figure}

	We propose a general method for distinguishing the possible causes of a sharp spectral feature. Consider a line of unknown cause---dark matter (DM), astrophysical, or detector---observed in the Milky Way (MW). Relative motion between source and observer leads to distinctive energy shifts as a function of line of sight (LOS) direction. Figure~\ref{fig:toon} illustrates this schematically. Because typical Galactic virial velocities are $\sim 10^{-3}c$, the Doppler shifts are only $\sim 0.1\%$. 
	
	A potential target for velocity spectroscopy is the 3.5-keV line recently observed in MW, M31, and galaxy cluster spectra~\cite{Bulbul:2014sua,Boyarsky:2014jta,Boyarsky:2014ska}. The line energy and flux can naturally be explained by sterile neutrino DM~\cite{Dodelson:1993je,Shi:1998km,Abazajian:2014gza,Abazajian:2001nj,Shaposhnikov:2006xi,Kusenko:2006rh,Merle:2013gea,Merle:2013wta,Patwardhan:2015kga,Venumadhav:2015pla} (or other candidates~\cite{Finkbeiner:2014sja,Higaki:2014zua,Lee:2014xua,Cicoli:2014bfa,Choi:2014tva,Frandsen:2014lfa,Dudas:2014ixa,Babu:2014pxa,Roland:2015yoa}). However, the significance of the line is disputed~\cite{Riemer-Sorensen:2014yda,Anderson:2014tza,Malyshev:2014xqa}, and it has been argued that it can be explained by astrophysical emission~\cite{Jeltema:2014qfa,Urban:2014yda}.

%
	With present detectors, velocity spectroscopy of this line is impossible. Excitingly, the Soft X-Ray Spectrometer (SXS) on Astro-H (launch date early 2016) has a goal energy resolution of $\sigma_{\rm{AH}} = 1.7 \, \rm{eV}$ (4 eV FWHM)~\cite{Takahashi:2012jn,2014SPIE.9144E..25T}, which is at the required $0.1\%$ scale. We show that if this goal resolution is achieved, Astro-H can identify the cause of the 3.5-keV line. We also discuss prospects if the performance is worse.
	
	We emphasize that the applicability of DM velocity spectroscopy is much more general. The purpose of this paper is to introduce a new concept to increase the power of DM searches and to spur innovation in detector design. We conclude by discussing several generalizations.	

\section{Usual DM Decay Signal}
\label{sec:Spectra}

	The differential intensity (flux per solid angle) from DM with mass $m_{\chi}$ and lifetime $\tau = 1/\Gamma$, decaying within the MW, is
\begin{equation}
\label{DiffI}
\frac{dI(\psi,E)}{dE}  = \frac{\Gamma}{4\pi m_\chi} R_{\odot} \rho_{\odot} \, \mathcal{J}(\psi)\frac{dN(E)}{dE} \, ,
\end{equation}
	where $R_{\odot} \simeq 8 \, \rm{kpc}$ and $\rho_{\odot} \simeq 0.4 \, \rm{GeV \, cm^{-3}}$~\cite{Catena:2009mf,2012ApJ...756...89B,2015arXiv150406324P} are the distance to the Galactic center (GC) and local DM density. (We neglect the cosmologically broadened extra-galactic signal, which contributes negligibly in Astro-H's narrow energy bins.) $\mathcal{J}(\psi)$ is the dimensionless, astrophysical J-factor defined by the LOS integral
\begin{equation}
\mathcal{J}(\psi) \equiv \frac{1}{R_{\odot}\, \rho_{\odot}} \int ds \, \rho_{\chi}(r[s,\psi]) \,,
\end{equation}
	where $\psi$ is the angle relative to the GC and is related to Galactic longitude and latitude via $\cos\psi = \cos l\cos b$. $dN(E)/dE$ is the photon spectrum.
 	
	The above treatment assumes that the astrophysical term, $ \mathcal{J}(\psi)$, and the photon spectrum, $ dN(E)/dE$, are separable. However, for detectors with energy resolution $\lesssim 0.1\%$, this approximation is not valid because relative velocities between source and observer, and therefore the spectral shape, vary along the LOS.
	
\vspace{-0.5 cm}
	
\section{Modified DM Spectrum}	
	We first account for how the signal is broadened by DM velocity dispersion and second for how it is shifted due to bulk relative motion. 
	
	We take the DM halo of the MW to be spherically symmetric, in steady state, and to have no appreciable rotation. The last is expected from angular momentum conservation, as the baryons from the proto-halo have collapsed significantly, while the DM has not; this is confirmed by simulations \cite{Bullock:2000ry,Vitvitska:2001vw}. Thus, $\langle \vec{v}_{\chi} \rangle =0$. 
	
	DM particles do have non-zero velocity dispersion, determined by the total gravitational potential of the halo~\cite{Binney:1987gd,Robertson:2009bh}. Assuming an isotropic velocity distribution ($\sigma_{v,r} =\sigma_{v,\phi} =\sigma_{v,\theta}$, so the total dispersion is $\sqrt{3}\sigma_{v,r}$), the radial velocity dispersion of DM is~\cite{Binney:1987gd}
\begin{equation}
	 \sigma_{v,r}^{2}(r)=\frac{G}{\rho_{\chi}(r)} \int_{r}^{R_{vir}} \! dr' \, \rho_\chi(r') \frac{M_{\rm{tot}}(r')}{r'^2} \, ,
\end{equation}
	where $M_{\rm{tot}}(r)$ is the total mass within a radius $r$. Typical values at $r \sim$~few kpc are $\sigma_{v,r} \simeq 125 \, \mathrm{km \, s^{-1}}$.
	
	To calculate $\sigma_{v,r}(r)$, we adopt the mass model of Ref.~\cite{Klypin:2001xu}, which fits a contracted DM and three-component baryon mass profile to MW rotation curve data; for more details see Supplemental Materials. The choice of mass model is not critical; kinematic results from other models agree within $\mathcal{O}(10\%)$ \cite{Catena:2009mf,McMillan:2011wd}.
	  
	The spectrum from a point along the LOS is the convolution of the intrinsic spectrum with the DM velocity distribution at that point. We assume a Maxwellian velocity distribution throughout the halo, which, at each point, yields a Gaussian distribution of the LOS velocity component. The modified spectrum from each point is
\begin{equation}
\frac{d\widetilde{N}(E,r[s,\psi])}{dE} = \int dE' \, \frac{dN(E')}{dE'} \, G(E-E';\sigma_{E'}) \, ,
\end{equation}
	where $G(E;\sigma_E)$ is a Gaussian of width $\sigma_{E} = (E/c)\sigma_{v_{\text{\tiny LOS}}}$. Based upon observations of the LOS velocity distribution of MW halo stars reported in \cite{Xue:2008se}, we take $\sigma_{v_{\text{\tiny LOS}}}(r) \simeq \sigma_{v,r}(r)$ which implies $\sigma_{E} =  (E/c) \, \sigma_{v,r}(r[s,\psi])$. 
	 
	The line shift follows from the LOS velocity, $v_{\text{\tiny LOS}} \equiv (\langle\vec{v}_{\chi}\rangle - \vec{v}_{\odot}) \cdot \hat{r}_{\text{\tiny LOS}}$, where positive $v_{\text{\tiny LOS}}$ indicates receding motion. For $v_{\text {\tiny LOS}} \ll c$, the resultant energy shift is $\delta E_{\rm{MW}}/E = {-v}_{\text{\tiny LOS}}/c$. 
	
	The Sun follows a roughly circular orbit about the GC in the direction toward positive Galactic longitude at a speed $v_{\odot} \simeq 220 \, \mathrm{km \, s^{-1}}$~\cite{Kerr:1986hz}. (Recent work suggests $v_{\odot} \gtrsim 240 \, \rm{km \, s^{-1}}$ \cite{Schonrich:2012qz,2012ApJ...759..131B}, which would strengthen our results.) The spectrum is therefore shifted by $\delta E_{\rm{MW}}(l,b)/E = +(v_{\odot}/c) \sin l\cos b$, which changes sign with $l$. We neglect the solar peculiar velocity as well as Earth and satellite motions, all of which are $\lesssim 10 \, \mathrm{km \, s^{-1}}$ \cite{McMillan:2009yr,Lee:2013xxa,Peter:2013aha}.   
	
	The final expression for the modified spectrum, including broadening and shifts, is therefore
	\begin{equation}
	\frac{d\mathcal{J}}{dE} = \frac{1}{R_{\odot}\rho_{\odot}} \int ds \, \rho_{\chi}(r[s,\psi]) \frac{d\widetilde{N}(E-\delta E_{\rm{MW}},r[s,\psi])}{dE} \,,
	\end{equation}
	so that Eq.~(\ref{DiffI}) is altered by $\mathcal{J}(\psi) \, dN(E)/dE \rightarrow d\mathcal{J}(\psi,E)/dE$. The observed signal, which is the convolution of $d\mathcal{J}/dE$ with the detector response, is nearly Gaussian and has an effective width $\sigma_{\rm{eff}}$.
	
\vspace{- 0.5 cm}	

\section{Modified Astrophysical Spectrum}
\label{Systems}

	The details are slightly different for astrophysical lines. 

	The widths of astrophysical lines are primarily determined by the mass of the emitting atom and by the gas temperature; turbulent broadening is negligible~\cite{Redfield:2004wb}. For potassium at $T = 2 \, \rm{keV}$, the intrinsic line width is $\sigma_{\rm{gas}}\simeq 0.8 \, \rm{eV}$, comparable to Astro-H's goal resolution, $\sigma_{\rm{AH}} \simeq 1.7 \, \rm{eV}$. The intrinsic width is weakly sensitive to the gas temperature and mass ($\propto \sqrt{T/m}$); any reasonable values of $T$ and $m$ give similar results.
	
	For the shift of an astrophysical signal, we must account for co-rotation within the MW disc. (While there is a non-rotating, gaseous halo at the outskirts of the MW, it is not hot enough to produce significant emission at 3.5 keV \cite{Dai:2011yn,Anderson:2011ih,Anderson:2014tza}). For simplicity, we assume all baryons follow circular orbits about the GC with speed $v_{\rm{circ}}(r) = \sqrt{G M_{\rm{tot}}(r)/r}$. With this circular speed and the hot gas distribution of Ref.~\cite{Ferriere:1998gm}, we compute the spectral shift by integrating the signal along the LOS with the contribution from each point weighted by the gas density. We call this fiducial model G2.
	
	Because the spatial and speed distributions of MW X-ray gas are uncertain, we compare to models  in Ref.~\cite{Kretschmer:2013naa} with smaller and larger line shifts. G1 is based on the distribution of free $e^-$~\cite{Cordes:2002wz} and the MW rotation curve~\cite{Sofue:2008wt}. G3 is based on the observed distribution of $^{26}$Al gamma rays \cite{Kretschmer:2013naa}. G1 and G2 are in good agreement with MW HI and CO data~\cite{Dame:2000sp,2008A&A487951K}. Peak LOS velocities for G1, G2, and G3 are $\simeq 50, 75,$ and $250 \, \rm{km \, s^{-1}}$. 	
	
\vspace{- 0.5 cm} 

\begin{figure}[t]
\centering
\includegraphics[width=\columnwidth]{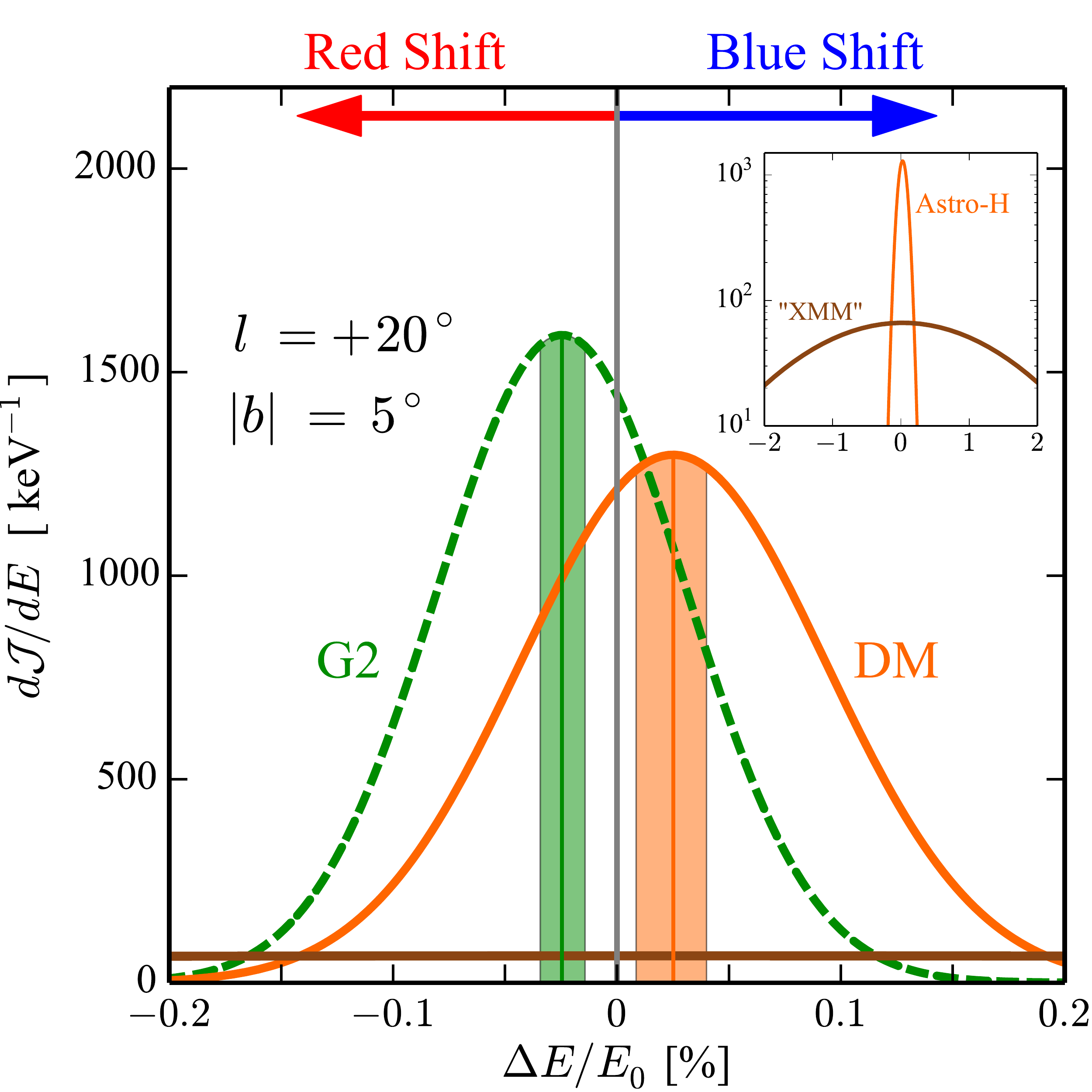}
\caption{Comparison of received spectra for DM and gas (G2). The emitted spectra are taken to have equal flux and to be centered at 3.5 keV before velocity effects. The line profiles include velocity dispersion and shift effects, as well as the energy resolution of Astro-H. Vertical bands indicate the 1-$\sigma$ centroid uncertainties after 2-Ms observations. For contrast, the brown line in the figure and inset shows the same signal if Astro-H had the energy resolution of XMM.}  
\label{fig:profiles}
\vspace{-0.25 cm}
\end{figure}


\begin{figure}[t]
\centering
\includegraphics[width=\columnwidth]{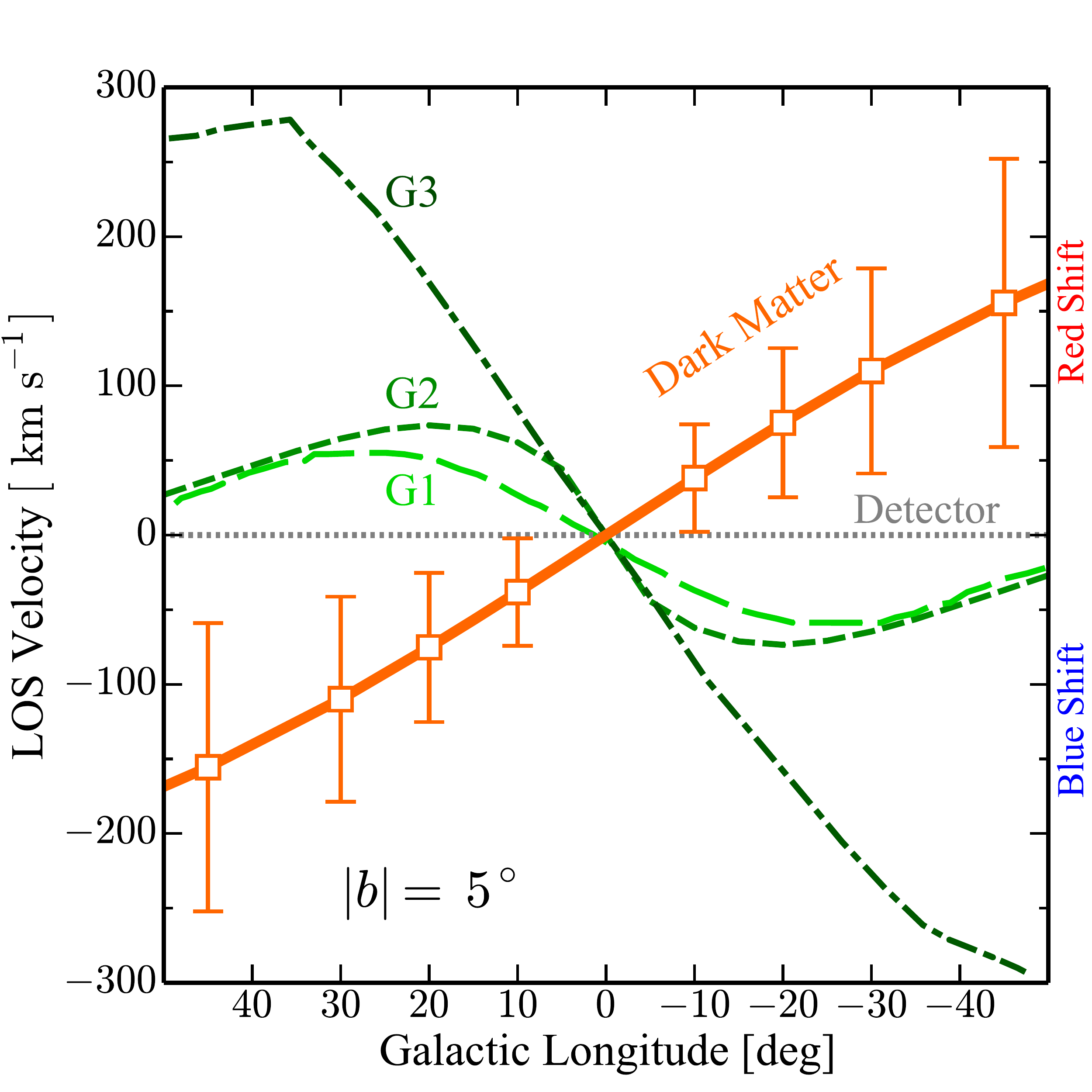}
\caption{LOS velocity for DM and various gas models (the realistic version of Fig.~\ref{fig:toon}).  Uncertainties are computed assuming 2-Ms Astro-H exposures on each point.}  
\label{fig:lvmap}
\vspace{-0.25 cm}
\end{figure}

\section{Line Flux Detection}
\label{sec:Prospects}
	
	One prerequisite to detecting a spectral shift is that the number of signal events be non-zero. Another is that the background fluctuations be small in comparison. Though Astro-H has a small field of view (FOV), its excellent energy resolution strongly suppresses backgrounds for a line signal, so that even a small number of signal events can be significant.
	
	Viewing directions $l \simeq 10^{\circ}-40^{\circ}$ have advantages. First, the balance between decreasing signal flux and increasing energy shift at large $l$ is optimized. Second, theoretical uncertainties are minimized, as the DM density profile at $r \gtrsim \rm{few \, kpc}$ is fixed by rotation curve data. Third, continuum astrophysical backgrounds are reduced; we reduce these further by going slightly off the Galactic plane, which minimally affects the DM signal.
		
	The expected signal intensity is calculated from Eq.~(\ref{DiffI}). For our DM example, this is
\begin{eqnarray}
I(\psi) &=& 1.2 \times 10^{-8} \, \mathrm{cm^{-2} \, s^{-1} \, arcmin^{-2}} \\
&& \times \left(\frac{\sin^{2}2\theta}{7 \times 10^{-11}}\right) \left(\frac{m_{\chi}}{7 \, \mathrm{keV}}\right)^{4} \left(\frac{\mathcal{J}(\psi)}{\mathcal{J}(l =20^{\circ}, |b| = 5^{\circ})}\right) \nonumber \, , 
\end{eqnarray}
 	where we have integrated over energy in the line profile, calculated $\mathcal{J}(l =20^{\circ}, |b| = 5^{\circ}) = 7.5$ using Ref.~\cite{Klypin:2001xu}, and taken the DM parameters from Ref.~\cite{Bulbul:2014sua}. For Astro-H, $\Omega_{\text{\tiny FOV}} = 9 \, \mathrm{arcmin}^{2}$ and $\mathrm{A_{eff}} = 200 \, \mathrm{cm}^{2}$~\cite{Takahashi:2012jn,2014SPIE.9144E..25T}, so the expected number of events is
\begin{equation}
N_{s}(\psi) \simeq 43 \, \left(\frac{\mathcal{J}(\psi)}{\mathcal{J}(l =20^{\circ}, |b| = 5^{\circ})}\right) \left(\frac{t}{2 \, \mathrm{Ms}}\right) \, .
\end{equation}
	This assumed exposure is large, but appropriate to the stakes (a potential discovery of DM) and the difficulties (the total exposure of XMM, Chandra, and Suzaku  used in the 3.5-keV analyses is $\gtrsim 40$ Ms~\cite{Bulbul:2014sua,Boyarsky:2014jta,Boyarsky:2014ska,Tamura:2014mta,Sekiya:2015jsa}). Furthermore, due to Astro-H's excellent energy resolution, all pointings in a substantial fraction of the sky will help test the 3.5-keV line.
	
	For continuum backgrounds, we consider only the contribution over the narrow energy range $\pm 2\sigma_{\rm{eff}}$ centered at 3.5 keV. (We do not need to include the tails of nearby astrophysical lines, as they will be well-resolved, unlike in XMM.)  One component of the background is due to the isotropic cosmic X-ray background (CXB)~\cite{Kushino:2002vk,Deluca:2003eu,Hickox:2005dz}. We conservatively adopt the total CXB flux (unresolved + resolved sources) $ E \, d\Phi_{\rm{CXB}}/dE = 9.2 \times 10^{-7} (\rm{E/keV})^{-0.4} \, \rm{cm^{-2} \, s^{-1} \, arcmin^{-2}}$~\cite{Hickox:2005dz}. Another background, due to hot gas in the MW, varies strongly with direction~\cite{Uchiyama:2012nw}. Finally, there are detector backgrounds due to intrinsic and induced radioactivities as well as cosmic-ray interactions; their intensity is expected to be comparable to that of the CXB~\cite{Kitayama:2014fda}. For $\psi(l=20^{\circ}, |b| = 5^{\circ})$, backgrounds contribute $N_b \simeq 5.2+5.4+5.4 = 16$ events per 2 Ms within the $\pm 2\sigma_{\rm{eff}} \simeq \pm 4.8 \, \rm{eV}$ band centered at 3.5 keV, compared to $N_s \simeq 41$. 
	
	We estimate the detection significance by the Poisson probability $P(n\geq 57 | \, \mu = 16)$, which corresponds to a one-sided Gaussian probability $> 7\sigma$. 
	
\vspace{-0.5 cm}
	
\section{Line Shift Detection}	

	Detecting a line shift depends on how well the centroid of the line profile is determined. Backgrounds decrease the precision, but, as above, the energy resolution of Astro-H plays a critical role. 
	
	When backgrounds are absent, the uncertainty on the centroid is $\sigma_{\rm{eff}}/\sqrt{N_s}$. When they are present, the uncertainty becomes $\delta E = C(R) \, \sigma_{\rm{eff}} / \sqrt{N_s}$, where C(R) is a correction factor and R is defined by the background to signal ratio. We calculate the optimal C(R) using the Cramer-Rao theorem \cite{PhysRev.88.775,Beacom:1998fj,James:2006zz}. For $\psi(l=20^{\circ}, |b| = 5^{\circ})$, $C(R) \simeq 1.6$, so that the uncertainty in the LOS velocity is $\delta_{v_{\text{\tiny LOS}}} \simeq 50 \, \rm{km \, s^{-1}}$.  
		
	Figure~\ref{fig:profiles} shows the  line profiles at $\psi(l=20^{\circ}, |b| = 5^{\circ})$ for a 3.5-keV emission line, due either to DM or gas. (A detector line would have zero shift). These profiles show how the energy spectra are shifted due to relative motion as well as broadened due to intrinsic dispersion and detector resolution. We show the uncertainties on the centroids, which are separated from each other and from zero in a 2-Ms exposure. With the energy resolution of XMM~\cite{Turner:2000jy} ($\sigma_{\rm{XMM}} \simeq 47 \, \rm{eV}$ vs.\ $\sigma_{\rm{AH}} \simeq 1.7 \, \rm{eV}$), the profiles are indistinguishable.
		
	Figure~\ref{fig:lvmap} shows how the expected shifts vary with Galactic longitude, along with their uncertainties, assuming 2-Ms observations for each point. We show the DM signal uncertainties; for an astrophysical line of the same flux, the uncertainties are comparable because the effective widths are comparable ($\sigma_{\rm{eff}}^{\rm{gas}} \simeq \, 160 \, \rm{km \, s^{-1}}, \, \sigma_{\rm{eff}}^{\text{\tiny DM}} \simeq \, 200 \, \rm{km \, s^{-1}}$); see Fig.~\ref{fig:profiles}. For a detector line with zero intrinsic width, the effective width is $ \sigma_{\rm{eff}}^{\rm{det}} \simeq 150 \,  \rm{km \, s^{-1}}$, approximately a factor of $\sqrt{2}$ less than $\sigma_{\rm{eff}}^{\text{\tiny DM}}$.
	
	For each point in Fig.~\ref{fig:lvmap}, it is easy to assess the probability that the expected DM signal could fluctuate to match that expected for an astrophysical or detector line, i.e., that a true DM signal could remain hidden. With two observations, at $l = \pm 20^{\circ}$, this scenario can be ruled out, relative to G2, at $\simeq 3.6 \sigma$. This establishes that this technique has interesting sensitivity. Once there is data, one can assess the probability that an astrophysical or detector line could mimic a DM signal (for the same flux, $ \delta_{v_{\text{\tiny LOS}}}^{\text{gas}} \simeq \delta_{v_{\text{\tiny LOS}}}^{\text{det}} \simeq \delta_{v_{\text{\tiny LOS}}}^{\text{\tiny DM}} / \sqrt{2} $).
	
	If the energy resolution is worse than the design goal, e.g., $\sigma_{\rm{AH}} \simeq$ 2.1, 2.5, or 3 eV, then the line shift significance is $\simeq$ 3.0, 2.4, or 1.9$\sigma$ (the line flux significance is always $> 5\sigma$).  This could be improved as $\sqrt{t}$ with more exposure (including non-dedicated pointings). We have not included the systematic uncertainty due to detector gain calibration, for which the goal is 0.4 eV~\cite{Kitayama:2014fda}. This can be mitigated by comparing the energies of nearby astrophysical lines, especially at opposite longitudes.

\vspace{-0.5 cm}

\section{Related Searches}
\label{sec: Gen}

	Astro-H may be able to resolve the intrinsic width of a MW DM line. This would provide the first information on the large-scale DM velocity distribution, which is sensitive to DM particle properties \cite{Rocha:2012jg} and to the presence of substructure~\cite{Ghigna:1999sn,2015ApJ...807...14L} (see Suppl. Mat.).

	The 3.5-keV line has been detected in M31. Due to the relative motion between the Sun and M31, DM or astrophysical lines from the center of M31 will have LOS shifts of $\simeq -300 \, \rm{km \, s^{-1}}$~\cite{Chemin:2009wd}. We estimate that this blue shift could be detected with $> 5\sigma$ significance, making this an attractive way to test detector causes. Due to M31's rotation, astrophysical lines are separated from DM lines by $\pm 200 \, \rm{km \, s^{-1}}$ around $\pm 1^{\circ}$, but, because the statistical uncertainties are large, they cannot be cleanly distinguished in 2 Ms; see Suppl. Mat. and Refs.~\cite{1997A&A...317..328S,Shirey:2000ar,Bender:2005rq,Bogdan:2008vs,Tamm:2012hw,Kaplinghat:2013xca,Ade:2015xua}. The LMC~\cite{vanderMarel:2002kq} may also be an attractive target.
	
	More speculatively, it may be possible to see the line in the extragalactic DM signal, if more astrophysical sources in the CXB are resolved, e.g., with eRosita \cite{Merloni:2012uf,Zandanel:2015xca}. Furthermore, because we move at $\simeq 400 \, \rm{km \, s^{-1}}$ with respect to the CMB, it may be possible to detect a dipole signature in DM line signal. Far-future observations may even detect a forest of sources in each LOS spectrum.

\vspace{-0.6 cm}


\section{Conclusions}
	
	Even for a supposedly smoking-gun signal, such as a line, it may be difficult to distinguish between DM, astrophysical, or detector causes. We have shown that detectors with energy resolution $\lesssim 0.1\%$ can break this degeneracy using velocity spectroscopy, which has minimal theoretical uncertainties. We emphasize that our main goal is to point out this new and robust method for testing DM signals, which can be applied to any sharp feature, such as an edge or box~\cite{Ibarra:2015tya,Boddy:2015efa}.
	
	To demonstrate the potential of this technique, we have shown that Astro-H will be able to test the origin of the 3.5-keV line.  In the future, other lines may be discovered. For lines at higher energy, the relative energy resolution of Astro-H improves. This unprecedented resolution will allow Astro-H to dramatically improve on existing sterile neutrino limits~\cite{Abazajian:2001vt,Boyarsky:2006fg,Watson:2006qb,Abazajian:2006jc,Abazajian:2006yn,Pullen:2006sy,Yuksel:2007xh,Boyarsky:2007ay,Boyarsky:2007ge,Loewenstein:2008yi,Boyarsky:2009ix,Abazajian:2011tk,Loewenstein:2012px,Jackson:2013pjq,Horiuchi:2013noa,Boddy:2014qxa,Ng:2015gfa,Figueroa-Feliciano:2015gwa,Riemer-Sorensen:2015kqa}. We encourage a dedicated study by the Astro-H Collaboration, once post-launch parameters are known, to give definitive answers on DM sensitivity over their full energy range. 

	We are encouraged by the expected $0.1 \%$ resolution of Astro-H in the range $0.3\!-\!12 \, \rm{keV}$, and the demonstrated $0.1 \%$ resolution of INTEGRAL-SPI in the range $20 \, \rm{keV}$ to $8 \, \rm{MeV}$ (including velocity spectroscopy of the 1.809-MeV line from $^{26}$Al~\cite{Kretschmer:2003ak,Diehl:2006cf,Kretschmer:2013naa}). Excitingly, the proposed X-ray mission ATHENA~\cite{Nandra:2013jka} and GeV gamma-ray mission HERD~\cite{Zhang:2014qga} have made achieving similar energy resolution a priority, which will improve existing limits~\cite{PhysRevLett.56.263,PhysRevD.37.3737,PhysRevD.40.3168,Bergstrom:1997fj,Hisano:2002fk,Gustafsson:2007pc,Mack:2008wu,2009PhRvD..80b3512B,Essig:2013goa,Ng:2013xha,Albert:2014hwa,TheFermi-LAT:2015gja}. We encourage other missions to pursue this aggressively.

\vspace{-0.5 cm}
	

\section*{Acknowledgments} 
	We are grateful to Yoshiyuki Inoue, Matthew Kistler, Greg Madejski, Phillip Mertsch, Annika Peter, and Randall Smith for discussions. EGS is supported by a Fowler Fellowship, KCYN and JFB by NSF grant PHY-1404311 to JFB, and RL by KIPAC.
			

\bibliographystyle{h-physrev}
\bibliography{dmvsbib}	

\begin{thebibliography}{100}

\bibitem{Bulbul:2014sua}
E.~Bulbul {\em et~al.},
\newblock Astrophys. J. {\bf 789}, 13 (2014), 1402.2301.

\bibitem{Boyarsky:2014jta}
A.~Boyarsky, O.~Ruchayskiy, D.~Iakubovskyi, and J.~Franse,
\newblock Phys. Rev. Lett. {\bf 113}, 251301 (2014), 1402.4119.

\bibitem{Riemer-Sorensen:2014yda}
S.~Riemer-Sorensen,
\newblock (2014), 1405.7943.

\bibitem{Boyarsky:2014ska}
A.~Boyarsky, J.~Franse, D.~Iakubovskyi, and O.~Ruchayskiy,
\newblock (2014), 1408.2503.

\bibitem{Anderson:2014tza}
M.~E. Anderson, E.~Churazov, and J.~N. Bregman,
\newblock (2014), 1408.4115.

\bibitem{Malyshev:2014xqa}
D.~Malyshev, A.~Neronov, and D.~Eckert,
\newblock Phys. Rev. {\bf D90}, 103506 (2014), 1408.3531.

\bibitem{Jeltema:2014qfa}
T.~E. Jeltema and S.~Profumo,
\newblock Mon. Not. Roy. Astron. Soc. {\bf 450}, 2143 (2015), 1408.1699.

\bibitem{Urban:2014yda}
O.~Urban {\em et~al.},
\newblock Mon. Not. Roy. Astron. Soc. {\bf 451}, 2447 (2015), 1411.0050.

\bibitem{Loewenstein:2009cm}
M.~Loewenstein and A.~Kusenko,
\newblock Astrophys. J. {\bf 714}, 652 (2010), 0912.0552.

\bibitem{Prokhorov:2010us}
D.~A. Prokhorov and J.~Silk,
\newblock Astrophys. J. {\bf 725}, L131 (2010), 1001.0215.

\bibitem{Weniger:2012tx}
C.~Weniger,
\newblock JCAP {\bf 1208}, 007 (2012), 1204.2797.

\bibitem{Finkbeiner:2012ez}
D.~P. Finkbeiner, M.~Su, and C.~Weniger,
\newblock JCAP {\bf 1301}, 029 (2013), 1209.4562.

\bibitem{2012arXiv1206.1616S}
M.~{Su} and D.~P. {Finkbeiner},
\newblock ArXiv e-prints  (2012), 1206.1616.

\bibitem{Aharonian:2012cs}
F.~Aharonian, D.~Khangulyan, and D.~Malyshev,
\newblock Astron. Astrophys. {\bf 547}, A114 (2012), 1207.0458.

\bibitem{Tempel:2012ey}
E.~Tempel, A.~Hektor, and M.~Raidal,
\newblock JCAP {\bf 1209}, 032 (2012), 1205.1045,
\newblock [Addendum: JCAP1211,A01(2012)].

\bibitem{Ackermann:2013uma}
[Fermi-LAT Collab.], M.~Ackermann {\em et~al.},
\newblock Phys. Rev. {\bf D88}, 082002 (2013), 1305.5597.

\bibitem{Weniger:2013tza}
C.~Weniger, M.~Su, D.~P. Finkbeiner, T.~Bringmann, and N.~Mirabal,
\newblock (2013), 1305.4710.

\bibitem{Ackermann:2015lka}
[Fermi-LAT Collab.], M.~Ackermann {\em et~al.},
\newblock Phys. Rev. {\bf D91}, 122002 (2015), 1506.00013.

\bibitem{Dodelson:1993je}
S.~Dodelson and L.~M. Widrow,
\newblock Phys. Rev. Lett. {\bf 72}, 17 (1994), hep-ph/9303287.

\bibitem{Shi:1998km}
X.-D. Shi and G.~M. Fuller,
\newblock Phys. Rev. Lett. {\bf 82}, 2832 (1999), astro-ph/9810076.

\bibitem{Abazajian:2014gza}
K.~N. Abazajian,
\newblock Phys. Rev. Lett. {\bf 112}, 161303 (2014), 1403.0954.

\bibitem{Abazajian:2001nj}
K.~Abazajian, G.~M. Fuller, and M.~Patel,
\newblock Phys. Rev. {\bf D64}, 023501 (2001), astro-ph/0101524.

\bibitem{Shaposhnikov:2006xi}
M.~Shaposhnikov and I.~Tkachev,
\newblock Phys. Lett. {\bf B639}, 414 (2006), hep-ph/0604236.

\bibitem{Kusenko:2006rh}
A.~Kusenko,
\newblock Phys. Rev. Lett. {\bf 97}, 241301 (2006), hep-ph/0609081.

\bibitem{Merle:2013gea}
A.~Merle,
\newblock Int.J.Mod.Phys. {\bf D22}, 1330020 (2013), 1302.2625.

\bibitem{Merle:2013wta}
A.~Merle, V.~Niro, and D.~Schmidt,
\newblock JCAP {\bf 1403}, 028 (2014), 1306.3996.

\bibitem{Patwardhan:2015kga}
A.~V. Patwardhan, G.~M. Fuller, C.~T. Kishimoto, and A.~Kusenko,
\newblock (2015), 1507.01977.

\bibitem{Venumadhav:2015pla}
T.~Venumadhav, F.-Y. Cyr-Racine, K.~N. Abazajian, and C.~M. Hirata,
\newblock (2015), 1507.06655.

\bibitem{Finkbeiner:2014sja}
D.~P. Finkbeiner and N.~Weiner,
\newblock (2014), 1402.6671.

\bibitem{Higaki:2014zua}
T.~Higaki, K.~S. Jeong, and F.~Takahashi,
\newblock Phys. Lett. {\bf B733}, 25 (2014), 1402.6965.

\bibitem{Lee:2014xua}
H.~M. Lee, S.~C. Park, and W.-I. Park,
\newblock Eur. Phys. J. {\bf C74}, 3062 (2014), 1403.0865.

\bibitem{Cicoli:2014bfa}
M.~Cicoli, J.~P. Conlon, M.~C.~D. Marsh, and M.~Rummel,
\newblock Phys. Rev. {\bf D90}, 023540 (2014), 1403.2370.

\bibitem{Choi:2014tva}
K.-Y. Choi and O.~Seto,
\newblock Phys. Lett. {\bf B735}, 92 (2014), 1403.1782.

\bibitem{Frandsen:2014lfa}
M.~T. Frandsen, F.~Sannino, I.~M. Shoemaker, and O.~Svendsen,
\newblock JCAP {\bf 1405}, 033 (2014), 1403.1570.

\bibitem{Dudas:2014ixa}
E.~Dudas, L.~Heurtier, and Y.~Mambrini,
\newblock Phys. Rev. {\bf D90}, 035002 (2014), 1404.1927.

\bibitem{Babu:2014pxa}
K.~S. Babu and R.~N. Mohapatra,
\newblock Phys. Rev. {\bf D89}, 115011 (2014), 1404.2220.

\bibitem{Roland:2015yoa}
S.~B. Roland, B.~Shakya, and J.~D. Wells,
\newblock (2015), 1506.08195.

\bibitem{Takahashi:2012jn}
T.~Takahashi {\em et~al.},
\newblock Proc. SPIE Int. Soc. Opt. Eng. {\bf 8443}, 1Z (2012), 1210.4378.

\bibitem{2014SPIE.9144E..25T}
T.~{Takahashi} {\em et~al.},
\newblock , SPIE Conference Series Vol. 9144, p.~25, 2014, 1412.1356.

\bibitem{Catena:2009mf}
R.~Catena and P.~Ullio,
\newblock JCAP {\bf 1008}, 004 (2010), 0907.0018.

\bibitem{2012ApJ...756...89B}
J.~{Bovy} and S.~{Tremaine},
\newblock Astrophys. J. {\bf 756}, 89 (2012), 1205.4033.

\bibitem{2015arXiv150406324P}
M.~{Pato}, F.~{Iocco}, and G.~{Bertone},
\newblock (2015), 1504.06324.

\bibitem{Bullock:2000ry}
J.~S. Bullock {\em et~al.},
\newblock Astrophys. J. {\bf 555}, 240 (2001), astro-ph/0011001.

\bibitem{Vitvitska:2001vw}
M.~Vitvitska {\em et~al.},
\newblock Astrophys. J. {\bf 581}, 799 (2002), astro-ph/0105349.

\bibitem{Binney:1987gd}
J.~Binney and S.~Tremaine,
\newblock {\em Galactic Dynamics} (Princeton University Press, 1987).

\bibitem{Robertson:2009bh}
B.~Robertson and A.~Zentner,
\newblock Phys. Rev. {\bf D79}, 083525 (2009), 0902.0362.

\bibitem{Klypin:2001xu}
A.~Klypin, H.~Zhao, and R.~S. Somerville,
\newblock Astrophys. J. {\bf 573}, 597 (2002), astro-ph/0110390.

\bibitem{McMillan:2011wd}
P.~J. McMillan,
\newblock Mon. Not. Roy. Astron. Soc. {\bf 414}, 2446 (2011), 1102.4340.

\bibitem{Xue:2008se}
[SDSS Collab.], X.~Xue {\em et~al.},
\newblock Astrophys. J. {\bf 684}, 1143 (2008), 0801.1232.

\bibitem{Kerr:1986hz}
F.~J. Kerr and D.~Lynden-Bell,
\newblock Mon. Not. Roy. Astron. Soc. {\bf 221}, 1023 (1986).

\bibitem{Schonrich:2012qz}
R.~Schonrich,
\newblock Mon. Not. Roy. Astron. Soc. {\bf 427}, 274 (2012), 1207.3079.

\bibitem{2012ApJ...759..131B}
J.~{Bovy} {\em et~al.},
\newblock Astrophys. J. {\bf 759}, 131 (2012), 1209.0759.

\bibitem{McMillan:2009yr}
P.~J. McMillan and J.~J. Binney,
\newblock Mon. Not. Roy. Astron. Soc. {\bf 402}, 934 (2010), 0907.4685.

\bibitem{Lee:2013xxa}
S.~K. Lee, M.~Lisanti, and B.~R. Safdi,
\newblock JCAP {\bf 1311}, 033 (2013), 1307.5323.

\bibitem{Peter:2013aha}
A.~H. Peter, V.~Gluscevic, A.~M. Green, B.~J. Kavanagh, and S.~K. Lee,
\newblock Phys. Dark Univ. {\bf 5-6}, 45 (2014), 1310.7039.

\bibitem{Redfield:2004wb}
S.~Redfield and J.~L. Linsky,
\newblock Astrophys. J. {\bf 613}, 1004 (2004), astro-ph/0406464.

\bibitem{Dai:2011yn}
X.~Dai, M.~E. Anderson, J.~N. Bregman, and J.~M. Miller,
\newblock Astrophys. J. {\bf 755}, 107 (2012), 1112.0324.

\bibitem{Anderson:2011ih}
M.~E. Anderson and J.~N. Bregman,
\newblock Astrophys. J. {\bf 737}, 22 (2011), 1105.4614.

\bibitem{Ferriere:1998gm}
K.~Ferriere,
\newblock Astrophys. J. {\bf 497}, 759 (1998).

\bibitem{Kretschmer:2013naa}
K.~Kretschmer {\em et~al.},
\newblock Astron. Astrophys. {\bf 559}, A99 (2013), 1309.4980.

\bibitem{Cordes:2002wz}
J.~M. Cordes and T.~Lazio,
\newblock (2002), astro-ph/0207156.

\bibitem{Sofue:2008wt}
Y.~Sofue, M.~Honma, and T.~Omodaka,
\newblock Publ. Astron. Soc. Jap. {\bf 61}, 227 (2009), 0811.0859.

\bibitem{Dame:2000sp}
T.~Dame, D.~Hartmann, and P.~Thaddeus,
\newblock Astrophys. J. {\bf 547}, 792 (2001), astro-ph/0009217.

\bibitem{2008A&A487951K}
P.~M.~W. {Kalberla} and L.~{Dedes},
\newblock Astron. Astrophys. {\bf 487}, 951 (2008), 0804.4831.

\bibitem{Tamura:2014mta}
T.~Tamura, R.~Iizuka, Y.~Maeda, K.~Mitsuda, and N.~Y. Yamasaki,
\newblock Publ. Astron. Soc. Jap. {\bf 67}, 23 (2015), 1412.1869.

\bibitem{Sekiya:2015jsa}
N.~Sekiya, N.~Y. Yamasaki, and K.~Mitsuda,
\newblock (2015), 1504.02826.

\bibitem{Kushino:2002vk}
A.~Kushino {\em et~al.},
\newblock Publ. Astron. Soc. Jap. {\bf 54}, 327 (2002), astro-ph/0204224.

\bibitem{Deluca:2003eu}
A.~De~Luca and S.~Molendi,
\newblock Astron. Astrophys. {\bf 419}, 837 (2004), astro-ph/0311538.

\bibitem{Hickox:2005dz}
R.~C. Hickox and M.~Markevitch,
\newblock Astrophys. J. {\bf 645}, 95 (2006), astro-ph/0512542.

\bibitem{Uchiyama:2012nw}
H.~Uchiyama, M.~Nobukawa, T.~G. Tsuru, and K.~Koyama,
\newblock Publ. Astron. Soc. Jap. {\bf 65}, 19 (2013), 1209.0067.

\bibitem{Kitayama:2014fda}
T.~Kitayama {\em et~al.},
\newblock (2014), 1412.1176.

\bibitem{PhysRev.88.775}
D.~E. Muller, H.~C. Hoyt, D.~J. Klein, and J.~W.~M. DuMond,
\newblock Phys. Rev. {\bf 88}, 775 (1952).

\bibitem{Beacom:1998fj}
J.~F. Beacom and P.~Vogel,
\newblock Phys. Rev. {\bf D60}, 033007 (1999), astro-ph/9811350.

\bibitem{James:2006zz}
F.~James,
\newblock {\em {Statistical Methods in Experimental Physics}} (North-Holland
  Publishing Co., 2006).

\bibitem{Turner:2000jy}
M.~J. Turner {\em et~al.},
\newblock Astron. Astrophys. {\bf 365}, L27 (2001), astro-ph/0011498.

\bibitem{Rocha:2012jg}
M.~Rocha {\em et~al.},
\newblock Mon. Not. Roy. Astron. Soc. {\bf 430}, 81 (2013), 1208.3025.

\bibitem{Ghigna:1999sn}
S.~Ghigna {\em et~al.},
\newblock Astrophys. J. {\bf 544}, 616 (2000), astro-ph/9910166.

\bibitem{2015ApJ...807...14L}
M.~{Lisanti}, D.~N. {Spergel}, and P.~{Madau},
\newblock Astrophys. J. {\bf 807}, 14 (2015), 1410.2243.

\bibitem{Chemin:2009wd}
L.~Chemin, C.~Carignan, and T.~Foster,
\newblock Astrophys. J. {\bf 705}, 1395 (2009), 0909.3846.

\bibitem{1997A&A...317..328S}
R.~{Supper} {\em et~al.},
\newblock Astron. Astrophys. {\bf 317}, 328 (1997).

\bibitem{Shirey:2000ar}
R.~Shirey {\em et~al.},
\newblock Astron. Astrophys. {\bf 365}, L195 (2001), astro-ph/0011244.

\bibitem{Bender:2005rq}
R.~Bender {\em et~al.},
\newblock Astrophys. J. {\bf 631}, 280 (2005), astro-ph/0509839.

\bibitem{Bogdan:2008vs}
A.~Bogdan and M.~Gilfanov,
\newblock Mon. Not. Roy. Astron. Soc. {\bf 388}, 56 (2008), 0803.0063.

\bibitem{Tamm:2012hw}
A.~Tamm, E.~Tempel, P.~Tenjes, O.~Tihhonova, and T.~Tuvikene,
\newblock Astron. Astrophys. {\bf 546}, A4 (2012), 1208.5712.

\bibitem{Kaplinghat:2013xca}
M.~Kaplinghat, R.~E. Keeley, T.~Linden, and H.-B. Yu,
\newblock Phys. Rev. Lett. {\bf 113}, 021302 (2014), 1311.6524.

\bibitem{Ade:2015xua}
[Planck Collab.], P.~A.~R. Ade {\em et~al.},
\newblock (2015), 1502.01589.

\bibitem{vanderMarel:2002kq}
R.~P. van~der Marel, D.~R. Alves, E.~Hardy, and N.~B. Suntzeff,
\newblock Astron. J. {\bf 124}, 2639 (2002), astro-ph/0205161.

\bibitem{Merloni:2012uf}
[eROSITA Collab.], A.~Merloni {\em et~al.},
\newblock (2012), 1209.3114.

\bibitem{Zandanel:2015xca}
F.~Zandanel, C.~Weniger, and S.~Ando,
\newblock (2015), 1505.07829.

\bibitem{Ibarra:2015tya}
A.~Ibarra, A.~S. Lamperstorfer, S.~L. Gehler, M.~Pato, and G.~Bertone,
\newblock (2015), 1503.06797.

\bibitem{Boddy:2015efa}
K.~K. Boddy and J.~Kumar,
\newblock (2015), 1504.04024.

\bibitem{Abazajian:2001vt}
K.~Abazajian, G.~M. Fuller, and W.~H. Tucker,
\newblock Astrophys. J. {\bf 562}, 593 (2001), astro-ph/0106002.

\bibitem{Boyarsky:2006fg}
A.~Boyarsky, A.~Neronov, O.~Ruchayskiy, M.~Shaposhnikov, and I.~Tkachev,
\newblock Phys. Rev. Lett. {\bf 97}, 261302 (2006), astro-ph/0603660.

\bibitem{Watson:2006qb}
C.~R. Watson, J.~F. Beacom, H.~Yuksel, and T.~P. Walker,
\newblock Phys. Rev. {\bf D74}, 033009 (2006), astro-ph/0605424.

\bibitem{Abazajian:2006jc}
K.~N. Abazajian, M.~Markevitch, S.~M. Koushiappas, and R.~C. Hickox,
\newblock Phys. Rev. {\bf D75}, 063511 (2007), astro-ph/0611144.

\bibitem{Abazajian:2006yn}
K.~Abazajian and S.~M. Koushiappas,
\newblock Phys. Rev. {\bf D74}, 023527 (2006), astro-ph/0605271.

\bibitem{Pullen:2006sy}
A.~R. Pullen, R.-R. Chary, and M.~Kamionkowski,
\newblock Phys. Rev. {\bf D76}, 063006 (2007), astro-ph/0610295.

\bibitem{Yuksel:2007xh}
H.~Yuksel, J.~F. Beacom, and C.~R. Watson,
\newblock Phys. Rev. Lett. {\bf 101}, 121301 (2008), 0706.4084.

\bibitem{Boyarsky:2007ay}
A.~Boyarsky, D.~Iakubovskyi, O.~Ruchayskiy, and V.~Savchenko,
\newblock Mon. Not. Roy. Astron. Soc. {\bf 387}, 1361 (2008), 0709.2301.

\bibitem{Boyarsky:2007ge}
A.~Boyarsky, D.~Malyshev, A.~Neronov, and O.~Ruchayskiy,
\newblock Mon. Not. Roy. Astron. Soc. {\bf 387}, 1345 (2008), 0710.4922.

\bibitem{Loewenstein:2008yi}
M.~Loewenstein, A.~Kusenko, and P.~L. Biermann,
\newblock Astrophys. J. {\bf 700}, 426 (2009), 0812.2710.

\bibitem{Boyarsky:2009ix}
A.~Boyarsky, O.~Ruchayskiy, and M.~Shaposhnikov,
\newblock Ann. Rev. Nucl. Part. Sci. {\bf 59}, 191 (2009), 0901.0011.

\bibitem{Abazajian:2011tk}
K.~N. Abazajian, P.~Agrawal, Z.~Chacko, and C.~Kilic,
\newblock Phys. Rev. {\bf D85}, 123543 (2012), 1111.2835.

\bibitem{Loewenstein:2012px}
M.~Loewenstein and A.~Kusenko,
\newblock Astrophys. J. {\bf 751}, 82 (2012), 1203.5229.

\bibitem{Jackson:2013pjq}
C.~Jackson, G.~Servant, G.~Shaughnessy, T.~M. Tait, and M.~Taoso,
\newblock JCAP {\bf 1307}, 021 (2013), 1302.1802.

\bibitem{Horiuchi:2013noa}
S.~Horiuchi {\em et~al.},
\newblock Phys. Rev. {\bf D89}, 025017 (2014), 1311.0282.

\bibitem{Boddy:2014qxa}
K.~K. Boddy, J.~L. Feng, M.~Kaplinghat, Y.~Shadmi, and T.~M.~P. Tait,
\newblock Phys. Rev. {\bf D90}, 095016 (2014), 1408.6532.

\bibitem{Ng:2015gfa}
K.~C.~Y. Ng, S.~Horiuchi, J.~M. Gaskins, M.~Smith, and R.~Preece,
\newblock (2015), 1504.04027.

\bibitem{Figueroa-Feliciano:2015gwa}
E.~Figueroa-Feliciano {\em et~al.},
\newblock (2015), 1506.05519.

\bibitem{Riemer-Sorensen:2015kqa}
S.~Riemer-Sorensen {\em et~al.},
\newblock (2015), 1507.01378.

\bibitem{Kretschmer:2003ak}
K.~Kretschmer, R.~Diehl, and D.~H. Hartmann,
\newblock Astron. Astrophys. {\bf 412}, L47 (2003), astro-ph/0311218.

\bibitem{Diehl:2006cf}
R.~Diehl {\em et~al.},
\newblock Nature {\bf 439}, 45 (2006), astro-ph/0601015.

\bibitem{Nandra:2013jka}
K.~Nandra {\em et~al.},
\newblock (2013), 1306.2307.

\bibitem{Zhang:2014qga}
[HERD Collab.], S.~Zhang,
\newblock (2014), 1407.4866.

\bibitem{PhysRevLett.56.263}
M.~Srednicki, S.~Theisen, and J.~Silk,
\newblock Phys. Rev. Lett. {\bf 56}, 263 (1986).

\bibitem{PhysRevD.37.3737}
L.~Bergstr\"om and H.~Snellman,
\newblock Phys. Rev. D {\bf 37}, 3737 (1988).

\bibitem{PhysRevD.40.3168}
A.~Bouquet, P.~Salati, and J.~Silk,
\newblock Phys. Rev. D {\bf 40}, 3168 (1989).

\bibitem{Bergstrom:1997fj}
L.~Bergstrom, P.~Ullio, and J.~H. Buckley,
\newblock Astropart. Phys. {\bf 9}, 137 (1998), astro-ph/9712318.

\bibitem{Hisano:2002fk}
J.~Hisano, S.~Matsumoto, and M.~M. Nojiri,
\newblock Phys. Rev. {\bf D67}, 075014 (2003), hep-ph/0212022.

\bibitem{Gustafsson:2007pc}
M.~Gustafsson, E.~Lundstrom, L.~Bergstrom, and J.~Edsjo,
\newblock Phys. Rev. Lett. {\bf 99}, 041301 (2007), astro-ph/0703512.

\bibitem{Mack:2008wu}
G.~D. Mack, T.~D. Jacques, J.~F. Beacom, N.~F. Bell, and H.~Yuksel,
\newblock Phys. Rev. {\bf D78}, 063542 (2008), 0803.0157.

\bibitem{2009PhRvD..80b3512B}
G.~{Bertone}, C.~B. {Jackson}, G.~{Shaughnessy}, T.~M.~P. {Tait}, and
  A.~{Vallinotto},
\newblock Phys. Rev. {\bf 80}, 023512 (2009), 0904.1442.

\bibitem{Essig:2013goa}
R.~Essig, E.~Kuflik, S.~D. McDermott, T.~Volansky, and K.~M. Zurek,
\newblock JHEP {\bf 11}, 193 (2013), 1309.4091.

\bibitem{Ng:2013xha}
K.~C.~Y. Ng {\em et~al.},
\newblock Phys. Rev. {\bf D89}, 083001 (2014), 1310.1915.

\bibitem{Albert:2014hwa}
[Fermi-LAT Collab.], A.~Albert {\em et~al.},
\newblock JCAP {\bf 1410}, 023 (2014), 1406.3430.

\bibitem{TheFermi-LAT:2015gja}
[Fermi-LAT Collab.],
\newblock Phys. Rev. {\bf D91}, 122002 (2015), 1506.00013.

\end{thebibliography}

\clearpage
\newpage

\onecolumngrid
\begin{center}
 \textbf{\large Supplemental Materials}
\end{center}
\vspace*{0.2cm}
\twocolumngrid

\section{Outline}

	We first briefly discuss the mass models and dispersion profiles used to derive the results presented in the main text. We then provide an expanded discussion of two additional applications of DM velocity spectroscopy, namely: probing the intrinsic DM dispersion profile using LOS observations and using velocity spectroscopy of M31 to test detector causes.
	
\vspace{-0.5 cm}	

\section{Radial Velocity Dispersion}

	To calculate the intrinsic broadening of a DM line, a galactic mass model must be adopted to determine the velocity dispersion profile. Below, we describe the mass models used in our analysis of the MW and M31.
	
\vspace{-0.5 cm}	
	
\subsection{Milky Way Mass Profile}
	
	We use model A1 of Ref.~\cite{Klypin:2001xu}, which utilizes a DM halo determined by adiabatically contracting an initial NFW profile in the presence of baryons.  We summarize key aspects of the model.
	
	Before contraction, an NFW profile with scale radius $r_s = 21.5$ kpc is assumed to coexist with three axisymmetric baryonic profiles roughly associated with the nucleus, bulge/bar, and disc of the galaxy. The total baryonic mass within a given radius is determined by the integration of the density profiles, with the addition of a central black hole of mass $m_{\rm{BH}}=2.6\times 10^6 M_{\odot}$. The enclosed baryonic mass is
\begin{equation}
	M_{\rm{b}}(r)=m_{\rm{BH}}+\int_{0}^{r}\int_{4\pi} dr' \, d\Omega \, \rho_{\rm{b}}(r') \, r'^2  \, .
\end{equation}

	The final DM profile is determined by contracting the initial NFW profile in the presence of this baryonic mass distribution. The baryonic profiles are adiabatically contracted under the assumption that spherical shells of matter do not cross and that the DM particles follow circular orbits. This deepens the potential well and causes the DM to contract. Angular momentum conservation then dictates the following equations:
\begin{eqnarray}
	 G \, [M_{\rm{b}}(r_f)+M_{\rm{dm}}(r_f)]\, r_f &=& G \, M_{\rm{halo}}(r_i) \, r_i\\
	 M_{\rm{halo}}(r_i) &=& M_{\rm{dm}}(r_f) \, \frac{(\Omega_{\rm{b}} + \Omega_{\rm{dm}})}{\Omega_{\rm{dm}}} \nonumber \, ,
\end{eqnarray}
	 where $M_{\rm{halo}}(r_i)$ is the halo mass before contraction and $\Omega_{\rm{dm}}$ and $\Omega_{\rm{b}}$ are the dark and baryonic matter densities, taken to be in the ratio $\Omega_{\rm{dm}}/(\Omega_{\rm{b}} + \Omega_{\rm{dm}}) = 0.9$; more recent observations give $\Omega_{\rm{dm}}/(\Omega_{\rm{b}} + \Omega_{\rm{dm}}) = 0.84$~\cite{Ade:2015xua}, which gives identical results.
	 
	 These equations are solved numerically to give a final radius, $r_f$, corresponding to a given initial radius, $r_i$. The contracted profile has a normalization $\rho_{\chi}(r = 8 \, \rm{kpc}) \simeq 0.4$ GeV $\rm{cm^{-3}}$. 
	 
	The combined baryonic and contracted DM profiles are integrated to give the total mass enclosed within a given radius, $M_{\rm{tot}}(r)$:
	\begin{equation}
		M_{\rm{tot}}(r) = M_b(r)+M_{\rm{dm}}(r) \, ,
	\label{Mtot}
	\end{equation}
	where $M_{\rm{dm}}(r)$ is the dark matter mass within a radius $r$. 
		
	The velocity dispersion is determined by the potential well of the galaxy, which is, in general, non-spherical. We approximate the true mass distribution by the spherically averaged mass profile given above. This approximation has little impact outside of $r \sim$~few kpc (where the DM becomes the dominant mass component), but greatly simplifies the calculation of the dispersion profile. Spherical symmetry allows for a simpler treatment of the Jeans equations~\cite{Binney:1987gd} and, together with equilibrium and an isotropic velocity distribution, yields the expression for the radial velocity dispersion given in the main text.
		
	This mass profile (Eq.~\ref{Mtot}) generates a rotation curve, $v_{\rm{circ}}(r) = \sqrt{G M_{\rm{tot}}(r)/r}$, which is in good agreement with observations and a dispersion profile which agrees with results of previous papers~\cite{Klypin:2001xu,Robertson:2009bh}.
		
\vspace{-0.5 cm}

\subsection{M31 Mass Profile}		

	We use the mass model of Ref.~\cite{Tamm:2012hw}. Generalized Einasto profiles (given below) are used to describe the baryonic components 
\begin{equation}
\rho_b(a) = \rho_c \, \exp\left(-d_N \, \left[\left(\frac{a}{a_c}\right)^{1/N}-1\right]\right) \, ,
\end{equation}
	with $\rho_c$, $d_N$, $a_c$, and N adjusted to match data. The baryonic mass model include five components (nucleus, bulge, disc, young disc, and stellar halo). Together with the adopted NFW DM profile, the measured M31 rotation curve is reproduced well~\cite{Tamm:2012hw}. 
	
	We also include a black hole of mass $m_{\rm{BH}}=3.5\times 10^7 M_{\odot}$~\cite{Klypin:2001xu}; more recent observations suggest a slightly larger mass or $1.4 \times 10^8 M_{\odot}$~\cite{Bender:2005rq}. The inclusion of a central black hole yields larger velocity dispersions at small radii ($\lesssim 10 \, \rm{pc}$), which increases the intrinsic width of DM lines arising from small angle LOS directions. However, because we focus on large angles ($l \simeq 10^{\circ}-40^{\circ}$ in the MW and $\psi \simeq 0.5^{\circ}- 1.5^{\circ}$ in M31), we do not probe the region affected by the black hole, so its effect is negligible; we verified that our results were unmodified by this addition. Dispersions in M31 are comparable to those in the MW, but are systematically higher because of its larger mass and concentration.
	
	Figure~\ref{fig:RDisp} shows the radial DM velocity dispersion profiles for the MW and M31. Vertical bands represent the range of radii that contribute $90\%$ to the signal along $\psi(l = 20^{\circ}, |b| = 5^{\circ})$ in the MW and $\psi = 1^{\circ}$ in M31.

\vspace{-0.5 cm}

\setcounter{figure}{0}
\renewcommand{\thefigure}{A\arabic{figure}}
\begin{figure}[t]
\centering
\includegraphics[width=\columnwidth]{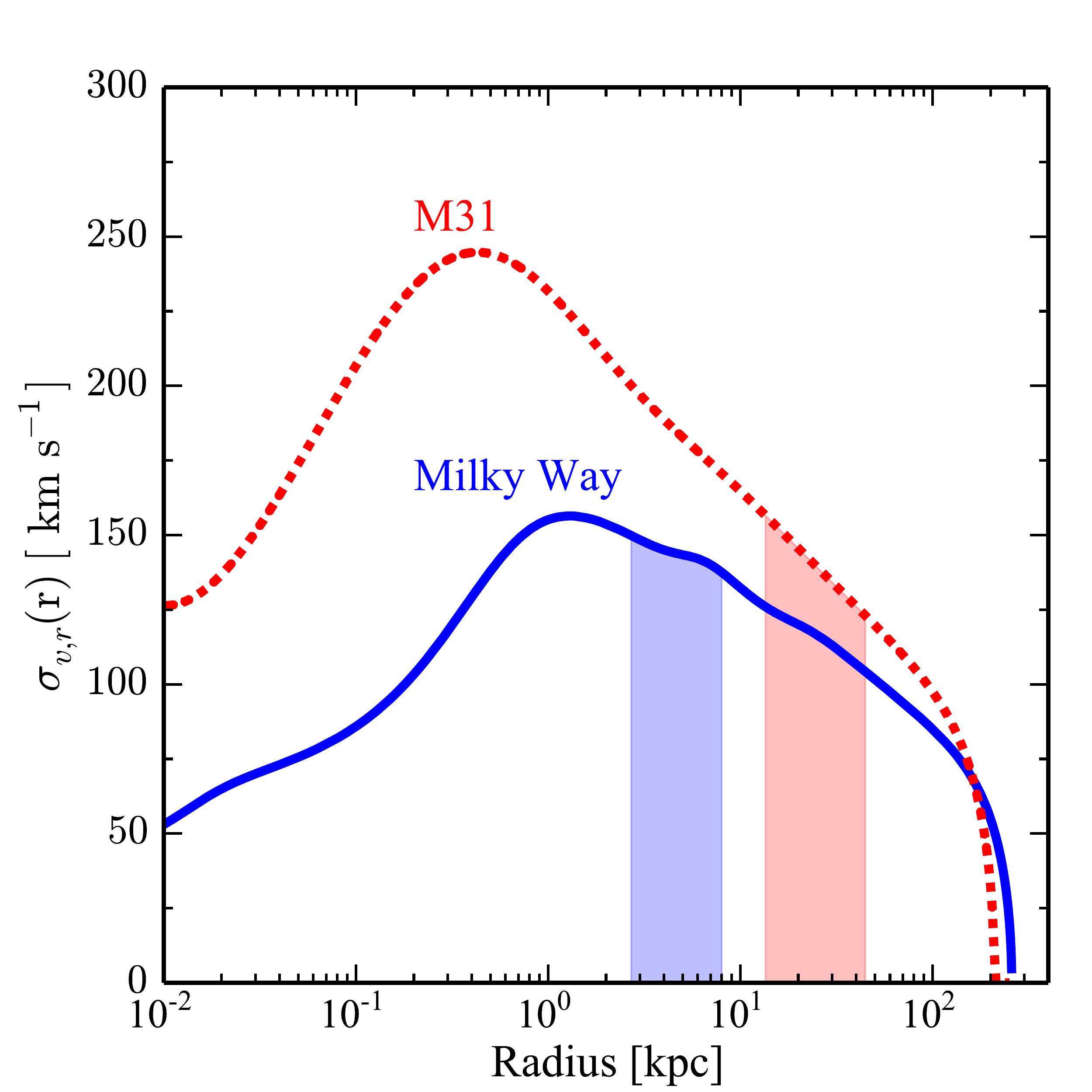}
\caption{Radial velocity dispersion profiles for the MW and M31. Shaded vertical bands indicate the range of radii that contribute $90\%$ of the signal along $\psi(l = 20^{\circ},|b| = 5^{\circ})$ in the MW and $\psi = 1^{\circ}$ in M31; the radius ranges for the other directions discussed in the text are similar. Note that the lower bounds of these ranges are the smallest $r$ probed by these directions.} 
\label{fig:RDisp}
\vspace{-0.25 cm}
\end{figure}

\begin{figure}[t]
\centering
\includegraphics[width=\columnwidth]{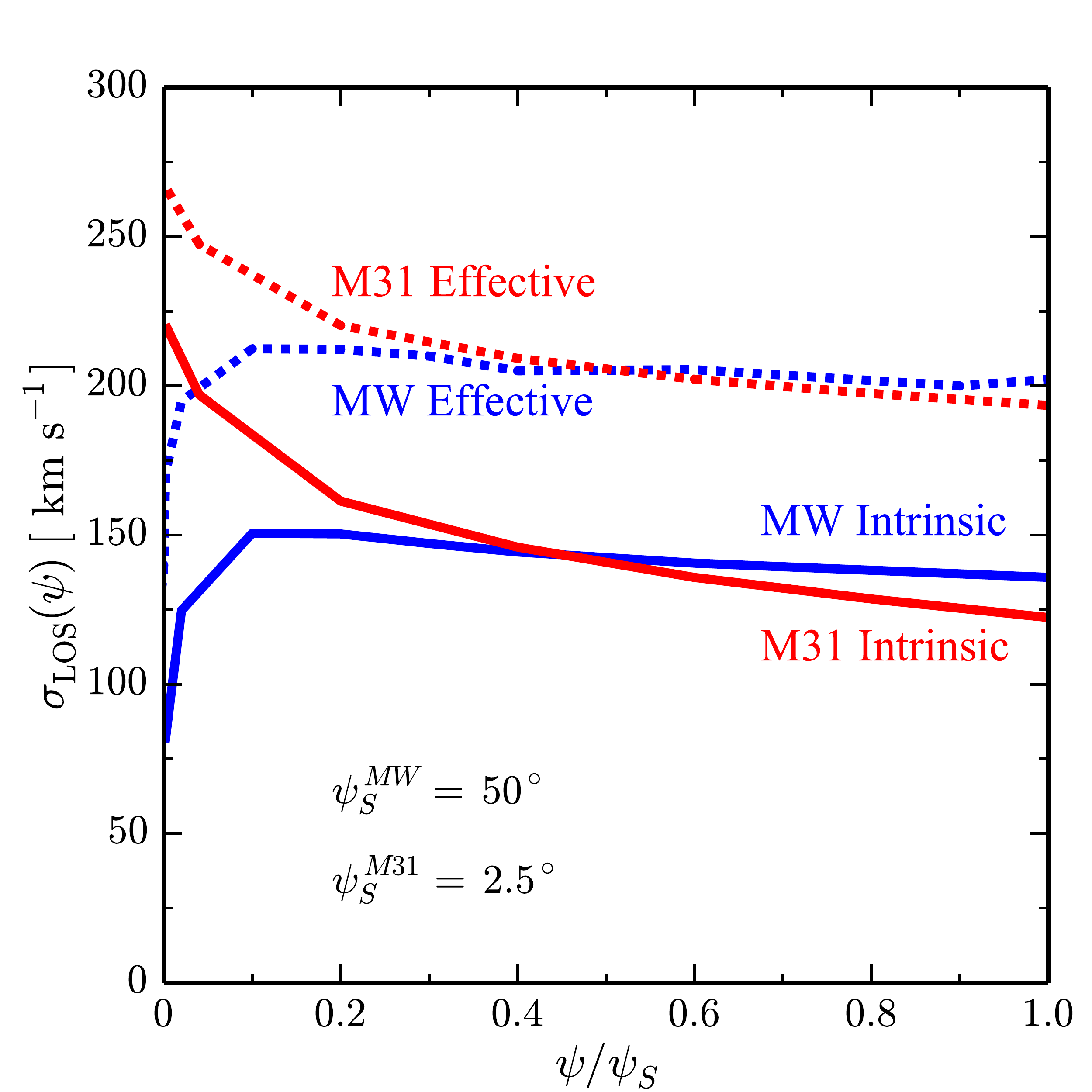}
\caption{Intrinsic ($\sigma_{\rm{DM}}$) and effective ($\sigma_{\rm{eff}}$) LOS velocity dispersion profiles for the MW and M31 as a function of $\psi/\psi_S$, the scaled angle relative to the center of each system. For the MW, $\psi_S = 50^{\circ}$, while for M31, $\psi_S = 2.5^{\circ}$; these scalings were chosen for display purposes. Intrinsic widths are determined by integrating the spectrum along the LOS using the radial velocity dispersion profiles given in the previous section. Effective widths include detector energy resolution. The increase in the LOS dispersion at small angles in M31 is due to the rising radial dispersions shown in Fig.~\ref{fig:RDisp}; for equally small (scaled) angles in the MW, the LOS dispersion decreases because only radii $< 1$ kpc, where the radial dispersion is decreasing, contribute.} 
\label{fig:LOSDisp}
\vspace{-0.25 cm}
\end{figure}

\begin{figure}[t]
\centering
\includegraphics[width=\columnwidth]{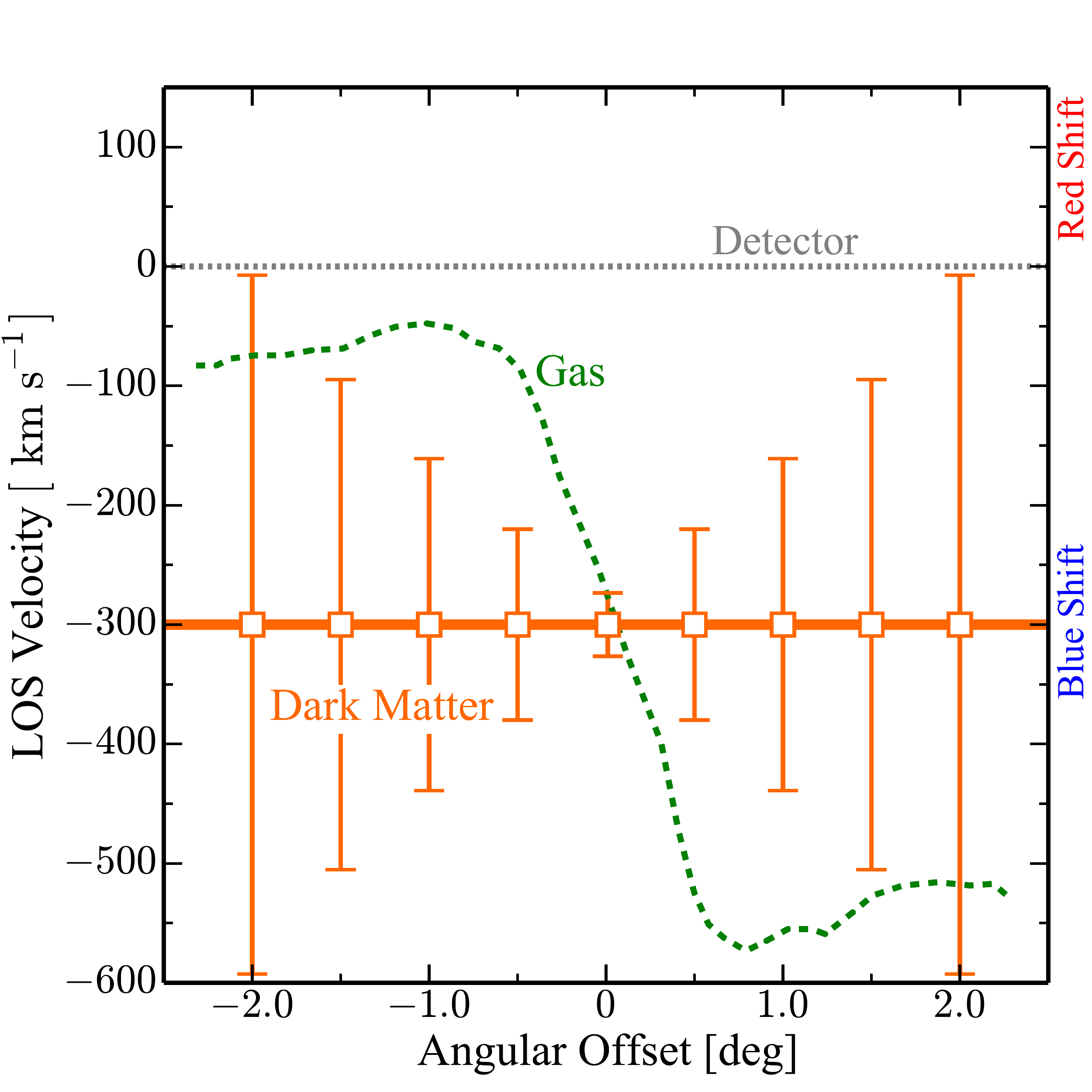}
\caption{LOS velocity profiles for DM and HI gas~\cite{Chemin:2009wd} in M31. DM error bars are calculated assuming 2-Ms exposures with Astro-H and only CXB and detector backgrounds.} 
\label{fig:M31LV}
\vspace{-0.25 cm}
\end{figure}

\section{LOS Velocity Dispersion}

	The velocity distribution of DM is of great interest both for the information it contains about the particle nature of DM and for its implications for direct and indirect detection experiments~\cite{Peter:2013aha}. For example, models of self-interacting DM (SIDM) predict higher velocity dispersions near the centers of DM halos. By measuring the LOS velocity dispersion, it may be possible to constrain SIDM interaction cross-sections, particularly in clusters where deviations between SIDM and CDM dispersions are large~\cite{Rocha:2012jg,Kaplinghat:2013xca}.  Additionally, because sub-halos generate smaller velocity dispersions, variations in line width along different LOS could help to constrain the size and distribution of DM substructure. 
	
	Because an observed DM signal will contain contributions from the entire LOS, and therefore a range of galactic radii, the full radial velocity dispersion cannot be probed directly. However, the observed LOS dispersion may still contain useful information. It is natural to ask how well Astro-H may be able to reconstruct the intrinsic DM LOS dispersion, given the observed signal.

	Figure~\ref{fig:LOSDisp} shows both the intrinsic and observed (assuming $\sigma_{\rm{AH}} \simeq 1.7 \, \rm{eV}$) LOS velocity width for a DM line in the MW and M31. Because the detector resolution is comparable to the intrinsic width, the detector response broadens the signal by a factor of $\simeq \sqrt{2}$. 
	
	In principle, if the energy resolution of Astro-H were known exactly, the intrinsic width of the DM line could be reconstructed precisely; assuming the signal and detector response are both Gaussian, the effective width is $\sigma_{\rm{eff}}^2 = \sigma_{\rm{AH}}^2 + \sigma_{\rm{DM}}^2$, so that $\sigma_{\rm{DM}}$ can be determined simply. 
	
	Of course, in practice, the resolution can never be known exactly. Assuming the goal uncertainty of 1 eV (the uncertainty is expected to be $\lesssim 2$ eV~\cite{Kitayama:2014fda}), we estimate that the intrinsic width of a 3.5-keV DM line can be reconstructed with an uncertainty of $\simeq 40 \, \rm{km \, s^{-1}}$; for higher energies the uncertainty in the width is smaller and scales as $E^{-1}$. See the Appendix of Ref.~\cite{Kitayama:2014fda} for more details regarding uncertainty in detector energy resolution and intrinsic line width reconstruction.
	
	More speculatively, using information about the strength of the signal along the LOS, it may be possible to construct a course-grained radial velocity dispersion profile from the LOS dispersion. For example, we see from the vertical bands in Fig.~\ref{fig:RDisp} that the range of radii that contributes to the $\psi(l = 20^{\circ}, |b| = 5^{\circ})$ signal is narrow and that the dispersion of these points is directly reflected in the intrinsic LOS dispersion shown in Fig.~\ref{fig:LOSDisp}. With additional pointings that probe different radii, it may be possible to constrain the radial dispersion profile using the measured line widths. This method would be most effective for small angles where the range of contributing radii is narrowest, although increased backgrounds would have to be overcome.
	
\vspace{-0.5 cm}

\section{Velocity Spectroscopy of M31}

	DM velocity spectroscopy can also be applied to a signal observed from M31. Relative motion between the Sun and M31 produces a DM LOS velocity shift of $\simeq -300 \, \rm{km \, s^{-1}}$ that is essentially independent of viewing angle. For astrophysical lines, one must also consider the rotation of the M31 disc. This produces an additional LOS velocity shift that varies strongly with viewing angle, separating the DM and astrophysical lines by $\pm \simeq 200 \, \rm{km \, s^{-1}}$ around $\pm 1^{\circ}$~\cite{Chemin:2009wd}. Detector lines are unshifted.
	
	The large differences in LOS velocities between DM, astrophysical, and detector lines make M31 a potentially powerful tool to probe the origin of spectral lines. However, large LOS velocities are not by themselves sufficient to distinguish between these three causes; it is also necessary that the uncertainty in the profile centroid be small in comparison to the expected centroid separations. 
	
	As discussed in the main text, the uncertainty in the centroid is given by $\delta E =  C(R) \, \sigma_{\rm{eff}} / \sqrt{N_s}$, where $\sigma_{\rm{eff}}$ is the observed line width, $N_s$ is the number of signal events, and C(R) is a correction factor that accounts for the presence of backgrounds. As can be seen in Fig.~\ref{fig:LOSDisp}, the observed widths of DM signals arising from M31 and the MW are expected to be quite similar. However, the number of DM signal events in M31 is considerably smaller, increasing the centroid uncertainty substantially. 
	
	Figure~\ref{fig:M31LV} shows the LOS velocities for DM, astrophysical and detector lines as a function of the angular offset $\psi$ from the center of M31. We show the error bars on a DM signal assuming 2-Ms observations and only CXB and detector backgrounds. Astrophysical X-ray emission in M31 is not well studied outside of $\sim 0.5^{\circ}$, but is expected to be small~\cite{1997A&A...317..328S,Shirey:2000ar,Bogdan:2008vs}. However, even without including this background, it is clear that the significance ($\propto \sqrt{t}$) with which DM and astrophysical signals can be differentiated is considerably smaller than for the MW. However, the large differences between DM, astrophysical and detector lines shifts could allow for cleaner separation of these causes, if uncertainties were reduced. If MW observations of a line suggest a DM origin, several Ms would be well spent on M31 observations. 
	
	Perhaps the greatest utility of observing M31 is in its power to test detector causes of a signal. This can be done most easily by looking directly at the center of M31. If the line is DM or astrophysical in nature, the signal strength should be strong and the centroid uncertainty correspondingly small, so that detector causes can be easily tested (Fig.~\ref{fig:M31LV}). Though we have shown error bars assuming 2-Ms observations, for this purpose, shorter exposures will clearly suffice.

\end{document}